\documentclass[conference]{IEEEtran}
\IEEEoverridecommandlockouts
\usepackage{amsmath,amssymb,amsfonts}
\usepackage{algorithmic}
\usepackage{graphicx}
\usepackage{float}
\usepackage{textcomp}
\usepackage{xcolor}
\usepackage{subfigure}

\def\BibTeX{{\rm B\kern-.05em{\sc i\kern-.025em b}\kern-.08em
    T\kern-.1667em\lower.7ex\hbox{E}\kern-.125emX}}
\begin{document}

\title{A Projection-Based Approach for Distributed Energy Resources Aggregation \\

\thanks{The work of Y. Wang and H. Zhong is supported by the National Key R\&D Program of China under Grant 2022YFB2403500.}
}

\author{\IEEEauthorblockN{Yiran Wang, Haiwang Zhong}
\IEEEauthorblockA{\textit{Department of Electrical Engineering, Tsinghua University}, Beijing, China \\
\textit{Sichuan Energy Internet Research Institute, Tsinghua University} \\
zhonghw@tsinghua.edu.cn}
\and
\IEEEauthorblockN{Guangchun Ruan}
\IEEEauthorblockA{\textit{MIT}, Boston, US}
}

\maketitle

\begin{abstract}
Aggregating distributed energy resources (DERs) is of great significance to improve the overall operational efficiency of smart grid. The aggregation model needs to consider various factors such as network constraints, operational constraints, and economic characteristics of the DERs. This paper constructs a multi-slot DER aggregation model that considers the above factors using feasible region projection approach, which achieved the protection of DERs data information and the elimination of internal variables. A system economic dispatch (ED) model is established for the operators to make full use of the DER clusters. We calculate the feasible regions with temporal coupling by extending the Progressive Vertex Enumeration (PVE) algorithm to high dimension by the Quickhull algorithm. Finally, an IEEE 39-bus distribution network is simulated with DERs to verify the effectiveness of the proposed model. Results show that the two-step ED derives the same results as the centralized ED.
\end{abstract}

\begin{IEEEkeywords}
aggregation, renewable energy, battery, optimal dispatch, feasible region, projection
\end{IEEEkeywords}

\section{Introduction}
An increasing number of distributed energy resources (DERs) in the power system brings significant challenges to the modern smart grid. It is crucial to make full use of these DERs to realize the potential benefits of aggregated flexibility~\cite{b16}. A DER aggregation model needs to reflect both the operational characteristics and cost characteristics of the DER clusters. Also, the aggregation model should consider the operational constraints of distribution networks in which the DERs locate at. For energy storage and other resources, temporal coupling constraints or binary states should be considered, even though this will largely increase the computational complexity. Up to now, it is often difficult to satisfy all of these conditions simultaneously.

The existing literature on DER aggregation often tend to formulate the feasible regions~\cite{b18}. Here the feasible region of each DER unit is constructed and then the Minkowski sum of all feasible regions is calculated in \cite{b1,b2,b3}. However, this approach does not consider the network operational constraints of distribution networks. As a result, the results of ED may lead to safety issues in the distribution network. A similar idea is applied in bidding decision where such regions could be derived by random sampling~\cite{b17}.

For system-level DER aggregation, network operational constraints must be considered. The existing literature often calculates the feasible region of DER clusters through projection. This can be implemented using the Fourier-Motzkin Elimination (FME) algorithm~\cite{b15}, which eliminates some variables in a set of linear inequalities by transforming some variables several times to project a set of constraints onto a specified variable. However, the computational complexity of the FME algorithm increases exponentially with the number of constraints and dimensionality, making it difficult to calculate high-dimensional feasible regions. In addition, there are also some sampling-based approximation methods, such as Monte Carlo method, Lagrangian relaxation method, which can be used to solve high-dimensional feasible region. These methods usually have higher computational efficiency, but lower calculation accuracy compared to projection-based methods.

While it is difficult to construct high-dimensional feasible regions, most research efforts are made in internal or external approximation algorithms for convex hulls. Reference \cite{b10} formulates the feasible region characterization problem as a maximum-minimum programming problem, which is transformed into a mixed integer programming problem later, and proposes an external approximation method based on the feasible cut. In \cite{b4}, the PVE (Progressive Vertex Enumeration) algorithm is proposed for characterizing the P-Q output feasible region of a VPP (Virtual Power Plant), which efficiently computes low-dimensional convex hulls. However, \cite{b4} only considers feasible region characterization for a single time slot and does not address the computational burden caused by temporal coupling. Reference~\cite{b6} proposes a method for computing the feasible power region of multi-slot tie-lines based on time slots clustering using PVE, by clustering the slots, the high-dimensional feasible region is divided into multiple low-dimensional feasible regions for calculation and finally, the Minkowski sum is calculated to obtain the feasible power region for multiple slots. Reference~\cite{b5} accurately calculates the aggregate feasible region (AFR) by analyzing all possible redundant constraints generated by FME and finding all non-redundant constraints as the feasible region of the model. Building upon \cite{b5}, the authors from \cite{b9} further formulate network constraints and temporal coupling in the mathematical expression for the entire flexibility of DERs.

The contributions of this paper are summarized as follows:
\begin{itemize}
\item A  feasible region calculation method is proposed for DER-rich distribution networks with network constraints, temporal coupling constraints, and cost characterization. % time slot
\item A two-step model is proposed to coordinate between DERs and distribution system operators, where the DER characteristics are captured by a set of dispatchable constraints for the smart grid operation.
\end {itemize}

In the remainder of this paper, Section \uppercase\expandafter{\romannumeral+2} establishes a model for distributed energy resources cluster systems, Section \uppercase\expandafter{\romannumeral+3} describes the feasible region calculation method based on projection, and Sections \uppercase\expandafter{\romannumeral+4} and \uppercase\expandafter{\romannumeral+5} respectively provide a case study and the conclusion.

\section{Problem Statement and Model}\label{SECPSM}

\subsection{Modeling of a Distributed Resource Cluster System}
The Distributed Energy Resources Cluster System (DCS) refers to a small-scale power system composed of DERs and the distribution network in which they are located. This system must satisfy the following network constraints of the distribution network:
\begin{equation}
P_{ij,t}^f = {g_{ij}}(v_{i,t}^2 - {v_{i,t}}{v_{j,t}}\cos {\theta _{ij,t}}) - {b_{ij,t}}{v_{i,t}}{v_{j,t}}\sin {\theta _{ij,t}}\label{eqn1}
\end{equation}
\begin{equation}
Q_{ij,t}^f =  - {b_{ij}}(v_{i,t}^2 - {v_{i,t}}{v_{j,t}}\cos {\theta _{ij,t}}) - {g_{ij}}{v_{i,t}}{v_{j,t}}\sin {\theta _{ij,t}},\label{eqn2}
\end{equation}
\begin{equation}
{(P_{ij,t}^f)^2} + {(Q_{ij,t}^f)^2} \le {({\bar S_{ij}})^2},\;{({P_0})^2} + {({Q_0})^2} \le {({\bar S_0})^2}\label{eqn3}
\end{equation}
\begin{equation}
P_{i,t}^{{\rm{DG}}} - P_{i,t}^{\rm{L}} = {g_{ii}}v_{i,t}^2 + \sum\limits_{j \in {{\cal L}_i}} {P_{ij,t}^f} ,\;i \in {{\cal L}_N}\label{eqn4}
\end{equation}
\begin{equation}
Q_{i,t}^{{\rm{DG}}} - Q_{i,t}^{\rm{L}} =  - {b_{ii}}v_{i,t}^2 + \sum\limits_{j \in {{\cal L}_i}} {Q_{ij,t}^f} ,\;i \in {{\cal L}_N}\label{eqn5}
\end{equation}
\begin{equation}
{\underline v}{_{i,t}} \le {v_{i,t}} \le {\overline{v}}{_{i,t}}\;i \in {{\cal L}_N},j \in {{\cal L}_i}\label{eqn6}
\end{equation}
where equations \eqref{eqn1} and \eqref{eqn2} are the active and reactive power flow equations of the distribution line $ij$ between bus $i$ and bus $j$, $P_{ij,t}^f$, $Q_{ij,t}^f$ represent the active and reactive power flows of the distribution line respectively, ${g_{ij}}$, ${b_{ij}}$ represent the series reactance and series conductance of the distribution circuit respectively, ${v_{i,t}}$, ${v_{j,t}}$  are the voltage magnitudes at bus $i$ and bus $j$ respectively, ${\theta _{ij,t}}$ represents the voltage phase angle difference between bus $i$ and bus $j$, ${{\cal L}_N}$ is the set of bus numbers, ${{\cal L}_i}$ representing the set of bus numbers that are directly connected to bus $i$ via distribution lines. \eqref{eqn3} represents the constraint on line/feeder capacity. \eqref{eqn4} and \eqref{eqn5} are the active and reactive power balance equations of the bus, where $P_{i,t}^{{\rm{DG}}}$ and $Q_{i,t}^{{\rm{DG}}}$ represent the active and reactive power output of distributed generation at bus $i$ respectively, $P_{i,t}^{\rm{L}}$ and $Q_{i,t}^{\rm{L}}$ represent the active and reactive loads at bus $i$ respectively. \eqref{eqn6} represents the constraint on voltage magnitude at buses.

This paper considers three types of DERs: distributed photovoltaic power generation (PV), distributed energy storage (ES), and flexible buildings (FB). The operational model for PV is shown as follows:
\begin{equation}
{\underline P}_t^{{\rm{PV}}} \le P_t^{{\rm{PV}}} \le {\overline{P}}_t^{{\rm{PV}}}\label{eqp1}
\end{equation}
\begin{equation}
{\left( {{\underline S}_t^{{\rm{PV}}}} \right)^2} \le {\left( {P_t^{{\rm{PV}}}} \right)^2} + {\left( {Q_t^{{\rm{PV}}}} \right)^2} \le {\left( {\overline{S}_t^{{\rm{PV}}}} \right)^2}\label{eqp2}
\end{equation}
\begin{equation}
P_t^{{\rm{PV}}}\tan \underline \phi _i^{{\rm{PV}}} \le Q_t^{{\rm{PV}}} \le P_t^{{\rm{PV}}}\tan \overline \phi _i^{{\rm{PV}}}\label{eqp3}
\end{equation}
where equations \eqref{eqp1}, \eqref{eqp2}, and \eqref{eqp3} represent the constraints on the active and reactive power output limits, capacity limit, and power factor limit of PV. ${\underline P}_t^{{\rm{PV}}}$ and ${\overline P}_t^{{\rm{PV}}}$ are the lower and upper limits of the active power output of PV, and ${\underline S}_t^{{\rm{PV}}}$ and ${\overline S}_t^{{\rm{PV}}}$ are the lower and upper limits of the apparent power output of distributed PV.

The operating model for energy storage is as follows:
\begin{equation}
\left\{ \begin{array}{l}
P_t^{{\rm{ESD}}} - P_t^{{\rm{ESD}}} = P_t^{{\rm{ES}}}\\
{E_{t{\rm{ + 1}}}} = {E_t} + P_t^{{\rm{ESC}}}{\eta _c} - \frac{{P_t^{{\rm{ESD}}}}}{{{\eta _d}}}\\
{E_T} = {E_{0{\rm{ }}}}\\
{\varphi _d} + {\varphi _c} = 1\\
\underline E  \le {E_t} \le \overline E \\
0 \le P_t^{\rm{ESD}} \le {{\overline P}^{{\rm{ESD}}}}{\varphi _d}\\
0 \le P_t^{\rm{ESC}} \le {{\overline P}^{{\rm{ESC}}}}{\varphi _c}\\
{\varphi _d},{\varphi _c} \in \left\{ {0,1} \right\}
\end{array} \right.
\end{equation}
where $P_t^{{\rm{ESD}}}$, $P_t^{{\rm{ESC}}}$, and $P_t^{{\rm{ES}}}$ represent the discharge power, charge power, and output power of energy storage at time $t$, respectively. $E_t$ is the energy storage capacity at time $t$, $\underline E$ and $\overline E$ are the lower and upper limits of energy storage capacity, $\varphi _c$ and $\varphi _d$ are the charging and discharging efficiencies of energy storage, ${\overline P}^{{\rm{ESD}}}$ and ${\underline P}^{{\rm{ESD}}}$ are the lower and upper limits of the discharge power, and ${\underline P}^{{\rm{ESC}}}$ and ${\overline P}^{{\rm{ESC}}}$ are the lower and upper limits of the charge power.

The flexible building operation model is as follows:
\begin{equation}
\left\{ \begin{array}{l}
{{\underline P}^{\rm{FB}}} \le P_t^{\rm{FB}} \le {{\overline P}^{\rm{FB}}}\\
\sum\limits_{t = 1}^T {P_t^{\rm{FB}}}  = P_s^{\rm{FB}}
\end{array} \right.
\end{equation}
where ${\underline P}^{\rm{FB}}$ represents the output power of flexible building at time $t$, and $P_s^{\rm{FB}}$ is the total power limit of flexible building over $T$ slots.

The total cost of DCS is as follows:
\begin{equation}
{C^{{\rm{DCS}}}} = \sum\limits_{t = 1}^T {\left( {{C^{{\rm{PV}}}}(P_t^{\rm{PV}}) + {C^{{\rm{ES}}}}(P_t^{\rm{ES}}) + {C^{{\rm{FB}}}}(P_t^{\rm{FB}})} \right)}
\end{equation}
where ${C^{{\rm{DCS}}}}$ is the total cost of DCS, ${C^{{\rm{PV}}}(.)}$ is the cost function of PV, $C^{{\rm{ES}}}(.)$ is the cost function of ES, ${C^{{\rm{FB}}}(.)}$ is the cost function of FB. This article set all the cost functions of DERs to be linear functions.

\subsection{Linearization and Relaxation of the DCS Model}
\subsubsection{Linearization of Network Constraints}
To obtain linear power flow equations, the linear network structure proposed in \cite{b12} is used. This method linearizes the network flow constraints and achieves satisfactory accuracy. For the power flow equations, with voltage phasor as the independent variable and ignoring network losses, the power flow equations can be approximated by a linear form:
\begin{equation}
P_{ij,t}^f = \frac{1}{2}{g_{ij}}\left( {{u_{i,t}} - {u_{j,t}}} \right) - {b_{ij}}\left( {{\theta _{i,t}} - {\theta _{j,t}}} \right)\\
\end{equation}
\begin{equation}
Q_{ij,t}^f =  - \frac{1}{2}{b_{ij}}\left( {{u_{i,t}} - {u_{j,t}}} \right) - {g_{ij}}\left( {{\theta _{i,t}} - {\theta _{j,t}}} \right)
\end{equation}

The capacity constraint \eqref{eqn3} is a convex quadratic function and can be piecewise linearized approximately (taking the line capacity constraint as an example):
\begin{equation}
P_{ij,t}^f\cos \frac{{2k\pi }}{N} + Q_{ij,t}^f\sin \frac{{2k\pi }}{N} \le {\bar S_{ij}}\cos \frac{\pi }{N},k \in \left[ {1,N} \right]
\end{equation}
where $N$ is the number of segments.

\subsubsection{Convex Relaxation of Energy Storage Constraints}
The operating constraints of energy storage contain 0-1 variables, which make the model strongly non-convex and cause the feasible region calculation by projection method to fail. Therefore, it is necessary to convexly relax the non-convex constraints. \cite{b7} considers the ED problem with energy storage and proposes an exact relaxation method for relaxing the non-convex model into a convex form under two sufficient conditions.

First, the discharge cost is set to be greater than or equal to the charging cost. In practice, this condition is often satisfied. If the energy storage is owned by the grid, the charging and discharging costs of the energy storage are both zero. If the energy storage is owned by a third party, in order to attract these owners to participate in the ED, the marginal compensation paid to them must cover their energy costs.

Second, it is ensured that the charging cost is always less than the marginal price. When the charging price is determined by the government or the grid company, the flexibility of charging may benefit the grid, so it can be reasonably expected that energy storage may receive rewards at a lower or even lower than marginal price, thus attracting more energy storage to participate in ED. Therefore, the second condition is easily satisfied in practice.

\section{Method of DCS aggregation and participation in ED}\label{SECMA}
\subsection{Projection-based feasible region calculation method}
This paper calculates the feasible region of DCS, which includes gate power for each slot and total cost, as the aggregation model of DCS for participation in power system ED. The calculation of the feasible region is based on variable space projection, and its theoretical foundation is as follows. For the DCS model established in \ref{SECPSM} after linearization and convex relaxation, its constraints can be expressed as follows:
\begin{equation}
Ax + By \le d\label{eqab}
\end{equation}
where $x$ is the variable to be eliminated, and $y$ is the variable retained after projection. In the process of DCS aggregation, $x$ usually represents internal decision variables of the DCS, such as bus voltage, branch power, and output of each DER. Meanwhile, $y$ usually represents the coordination variables between the DCS and the grid, such as the gate power and cost.

The feasible region constructed by formula\eqref{eqab}is a polyhedron in space ${\Re^{{N_x}}} \times {\Re^{{N_y}}}$, which can be represented as:
\begin{equation}
{\rm{\Omega}} = \left\{ {\left( {x,y} \right) \in {\Re^{{N_x}}} \times {\Re^{{N_y}}}:Ax + By \le b} \right\}
\end{equation}\label{eq8}

The feasible region of projecting ${\Omega}$ onto a subspace is as follows:
\begin{equation}
{\rm{\Xi }} = \left\{ {y \in {\Re^{{N_y}}}:\exists x \in {\Re^{{N_x}}},{\mathop{\rm s}\nolimits} .t{\rm{.}}\left( {x,y} \right) \in {\rm{\Omega }}} \right\}
\end{equation}\label{eq9}

The feasible region of ${\rm{\Xi }}$ as a coordination variable $y$ can be used as the aggregation model of the DCS to participate in power system ED. This paper utilizes the PVE algorithm proposed in \cite{b8} for DCS feasible region calculation based on subspace projection.

\subsection{PVE Algorithm}
PVE algorithm is an efficient projection algorithm for vertex-based convex hulls. This algorithm constructs the convex hull from initial vertices, uses the outward normal direction of each surface of the convex hull as the search direction to identify new vertices on the convex hull, and adds new vertices to the convex hull in each iteration until it meets the precision requirements. This method is an inner-approximation method of the convex hull, and its efficiency is significantly improved compared to FME. By utilizing the PVE algorithm, we can obtain either the vertices or the hyperplane inequalities that represent the feasible region of the convex hull. The vertex search problem is equivalent to solving the following optimization problem:
\begin{equation}
\begin{array}{l}
\mathop {\max }\limits_x \;h = {\eta ^ \top }x\\
{\rm{s}}{\rm{.t}}{\rm{.}}\;\;\;\;(y,x) \in \Omega .
\end{array}
\end{equation}

The key to vertex identification is to select appropriate coefficients for the objective function, which represents the direction of searching for vertices. The PVE algorithm follows the following steps:

(1) Initialize vertices. To construct an $N$-dimensional convex hull, at least $N$+1 vertices are required. Initialize $N$+1 $N$-dimensional vectors and perform a search using the equation $\eta$ to obtain an initial set of $N$+1 vertices.

(2) Search for new vertices. Calculate the equation of the convex hull hyperplane based on the initial set of vertices, and search for new vertices along the outer normal direction vector of the hyperplane. Add the new vertex to the vertex set.

(3) Termination. Calculate the expansion amount of the convex hull and repeat step (2) until the expansion amount is less than $\epsilon$.

The PVE algorithm proposed above is only applicable to feasible region calculations with a single time slot. This paper extend the PVE algorithm to compute high-dimensional convex hulls for multiple time profiles. The calculation of the convex hull hyperplane equation in step (2) depends on the Polyhedron convex hull calculation function in the Multi-Parametric Toolbox\cite{b13}, which is the main obstacle that limits the computation dimensionality of the PVE algorithm. This paper replaces this function with a convex hull calculation method based on Quickhull algorithm\cite{b11}, achieving the extension of the PVE algorithm from single time slot to multiple time slots.

The process of calculating the convex hull based on the Quickhull algorithm is as follows:

(1) Calculate the index set of points ${\Theta ^M}$ that make up the surface of the convex hull in the set of points $V$ using the Quickhull algorithm. There are $M$ sets of point indices corresponding to $\omega$ surfaces of the convex hull.

(2) For the $i$th set of point indices $\Theta _i^M$, calculate the outward normal vector of the $i$th surface of the convex hull using ${\alpha _i} = {\left( {V\left( {\Theta _i^M} \right)} \right)^{ - 1}} \cdot {\bf{1}}$.

(3) Calculate the centroid of the convex hull ${V_0}$ and the $\sigma _i = \alpha_i^{\rm{T}}{V_0} - 1$. If ${\sigma _i} \ge 0$, ${V_0}$ is inside the hyperplane $\alpha _i^{\rm{T}}x - 1 = 0$, and set ${b_i} = 1$; if ${\sigma _i} < 0$, ${V_0}$ is outside the hyperplane , and set ${b_i} =  - 1$, ${\alpha _i} =  - {\left( {V\left( {\Theta _i^M} \right)} \right)^{ - 1}} \cdot {\bf{1}}$.

(4) Repeat steps (2) and (3) until obtaining a set of vectors $A = \left[ {{\alpha _1},{\alpha _2} \ldots {\alpha _\omega }} \right]$ that form the convex hull of the given point set $V$. The hyperplane equation of the convex hull can be expressed as ${A^{\rm{T}}}x \le b$.

After improving the PVE algorithm, high-dimensional vertex search for the convex hull can be achieved, which can be used for computing feasible region in multiple time slots.

\subsection{DCS-Grid ED Model}\label{SECDOM}
The high-dimensional vertex set $V$ obtained in Section \ref{SECMA} is the feasible region of the DCS gate power and total cost, which can be represented by the following convex combinations:
\[{{\rm{\Xi }}^{DCS}} = \left\{ {{{\left( {{P^{{\rm{DCS}}}},{C^{{\rm{DCS}}}}} \right)}^{\rm{T}}} = \sum\limits_i^n {{\lambda _i}V} ,\;\sum\limits_i^n {{\lambda _i}}  = 1,\;{\lambda _i} \ge 0} \right\}\]
where ${{\left( {{P^{{\rm{DCS}}}},{C^{{\rm{DCS}}}}} \right)}^{\rm{T}}}$ is a \textit{T}-dimensional vector that represents the DCS gate power in each time slot, with positive direction as input, ${C^{{\rm{DCS}}}}$ is the total cost of DCS and is a convex combination of \textit{V}, \textit{n} is the number of vertices in \textit{V}.

After DCS is integrated into the smart grid, it forms a DCS-Grid system, and the ED model of this system is as follows:
\begin{equation}
\begin{array}{l}
\quad \;\;\;\min \;\;{C^{{\rm{DCS}}}} + \sum\limits_{t = 1}^T {C_t^{{\rm{CS}}}} \\
{\rm{s}}{\rm{.t}}{\rm{.}}\;\;\left\{ \begin{array}{l}
\left( {{P^{{\rm{DCS}}}},{C^{{\rm{DCS}}}}} \right) \in {{\rm{\Xi }}^{{\rm{DCS}}}}\\
P_t^{{\rm{PM}}} = P_t^{\rm{L}} + P_t^{{\rm{DCS}}}\\
{\underline P ^{{\rm{PM}}}} \le P_t^{{\rm{PM}}} \le {\overline P ^{{\rm{PM}}}}\\
 - {\delta ^{{\rm{PM}}}}\Delta t \le P_t^{{\rm{PM}}} - P_{t - 1}^{{\rm{PM}}} \le {\delta ^{{\rm{PM}}}}\Delta t\\
C_t^{{\rm{CS}}} = {C^{{\rm{PM}}}}\left( {P_t^{{\rm{PM}}}} \right)
\end{array} \right.
\end{array}
\end{equation}
where $C_t^{{\rm{CS}}}$ is the cost of the grid at time $t$, $P_t^{\rm{L}}$ is the load of the grid at time $t$, $P_t^{{\rm{PM}}}$ is the output power of the thermal power unit at time $t$, ${\underline P ^{{\rm{PM}}}}$ and ${\overline P ^{{\rm{PM}}}}$ are the lower and upper limits of the active power output of thermal power unit, ${\delta ^{{\rm{PM}}}}$ is the ramping limit of the thermal power unit, $\Delta t$ is the duration of each time slot and ${C^{{\rm{PM}}}}$ is the cost functions of the thermal power unit. Considering that most distributed generations are renewable energy sources, the cost of thermal power unit is set to be higher than the cost of distributed generations.

\section{Case Study}
\subsection{Basic Setup}
\begin{figure}[htbp]
  \centering
  \includegraphics[width=0.85\linewidth]{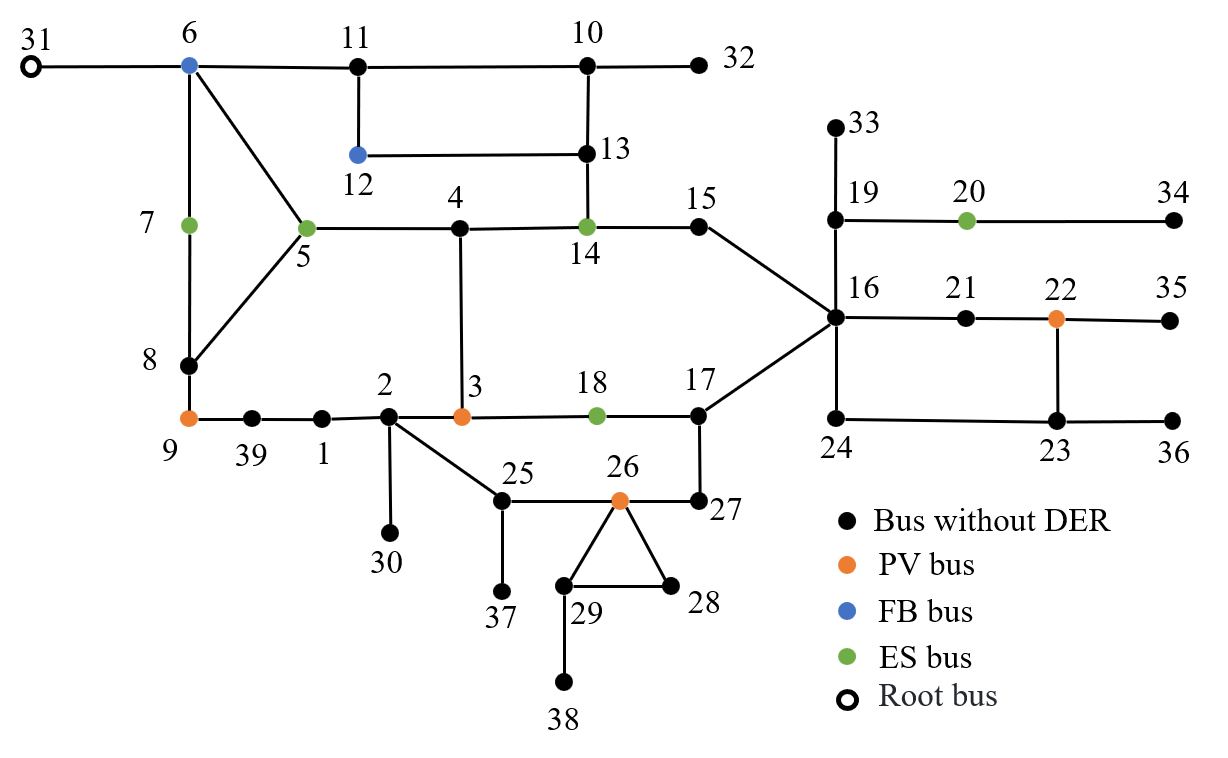}
  \caption{Modified IEEE 39-bus distribution system}
  \label{figSYS}
\end{figure}

This paper performs ED during six time slots from 12:00 to 17:00 and the duration of each time slot is one hour. The system has 4 PV connected to buses 3, 9, 22, and 26, 5 ES connected to buses 5, 7, 14, 18,and 20, and 2 FB connected to buses 6 and 12, as shown in the figure \ref{figSYS}. The boundary operating conditions include power loads and PV outputs. Hourly data is assumed based on the 24-hour load and solar irradiance data collected from an actual distribution network. The DCS is connected to a grid consisting of thermal power unit at bus 31, without considering network constraints.

Numerical simulations are performed on a computing environment with Intel(R) Core(TM) i9-13900K CPUs running at 3.00 GHz and with 64-GB RAM. All programming is implemented in Matlab 2022b.

\subsection{Simulation Results}
To calculate the feasible region of the DCS, the PVE algorithm is used based on linearized power flow equations and distribution network constraints, following the calculation steps described in Section \ref{SECMA} of this paper. The feasible region is then used as an aggregated model for DERs to participate in ED (S2), and compared with the results of centralized ED (S1).

\begin{figure}[htbp]
\centering
\subfigure[1st iteration]{
\includegraphics[width=0.43\linewidth]{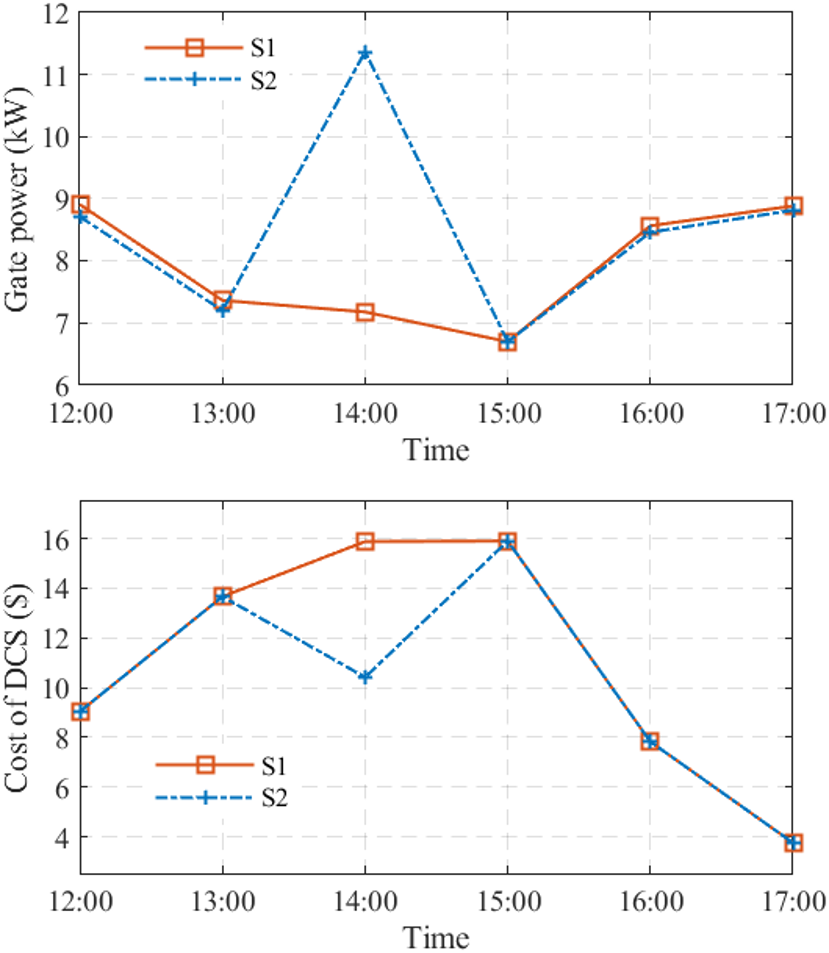}
\label{figS1S2a}
}
\quad
\subfigure[2nd iteration]{
\includegraphics[width=0.43\linewidth]{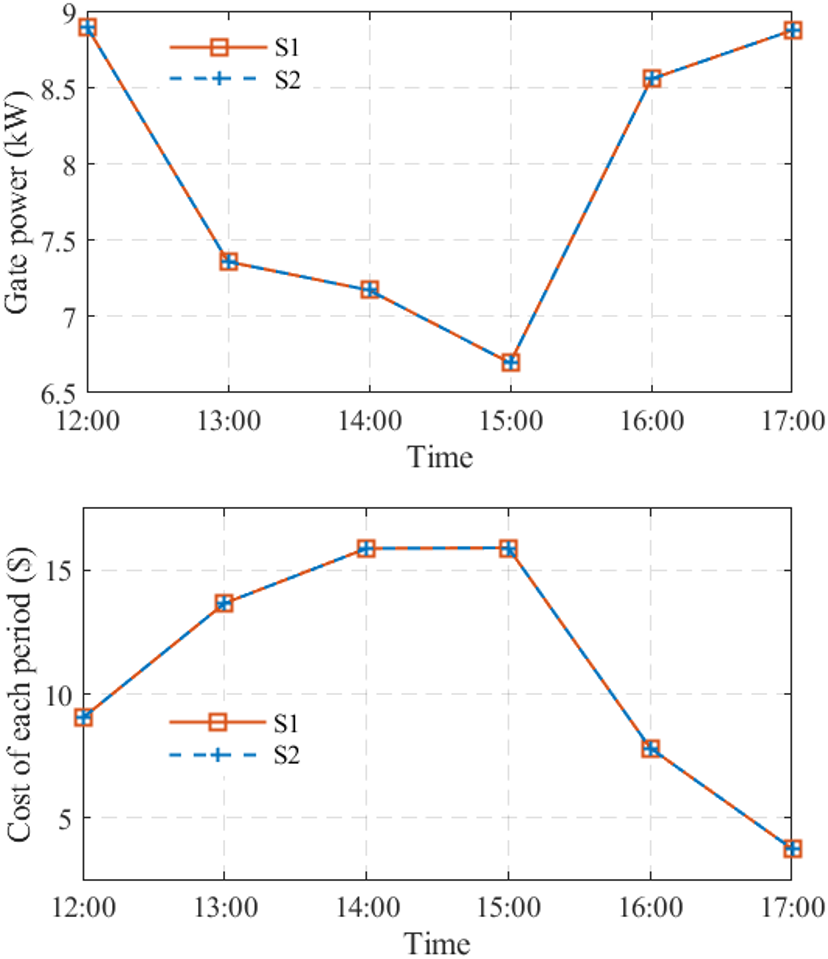}
\label{figS1S2b}
}
\caption{Gate power and cost for each time slot in each iteration}
\label{figS1S2}
\end{figure}

The feasible region of the DCS is a high-dimensional convex set with a dimension of 7 (including gate power for 6 time slots and the total cost) and 24046 vertices. Data during iterations are shown in the table \ref{tableS1S2}. The comparison of gate power and cost after using methods S1 and S2 for ED is shown in the figure \ref{figS1S2}. The figure \ref{figS1S2a} shows the result of ED using the convex hull containing 139 points obtained in the first iteration as the aggregation model. It can be seen that the ED results of S2 are significantly different from those of S1. From Figure 2, it can be seen from figure \ref{figS1S2b} that the ED results of the second iteration using the convex hull as the aggregation model for ED are basically the same as those of centralized ED. And the total system cost under both methods is 2607.07 USD. This indicates that the proposed method of aggregating can well reflect the boundary power and cost characteristics of the DERs cluster.

\begin{figure}[htbp]
\centering
\includegraphics[width=0.9\linewidth]{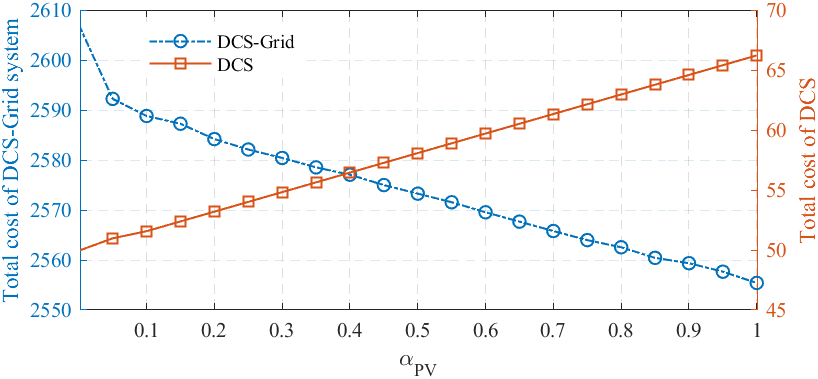}
\caption{Total cost of DCS-Grid system and DCS with different PV capacities}
\label{figPVcost}
\end{figure}

To illustrate the improvement of the overall benefits of the grid with the addition of DERs, we set the variable ${\alpha _{{\rm{PV}}}}$ representing the PV capacity of DCS, where ${\alpha _{{\rm{PV}}}}$ = 0 represents no PV in DCS, and ${\alpha _{{\rm{PV}}}}$ = 1 represents the PV capacity reaching the capacity set in the case in \ref{SECDOM}. As can be seen from the figure \ref{figPVcost}, as the PV capacity increases, though the cost of the DCS is increasing, the total cost of the DCS-Grid system gradually decreases, indicating that the participation of DERs in smart grid helps to increase the overall system benefits.
\begin{table}[htbp]
\centering
\caption{Data during Iterations}
\label{tableS1S2}
\renewcommand{\arraystretch}{1.5}
\setlength{\tabcolsep}{6pt}
\begin{tabular}{ccc}
\hline
\multicolumn{1}{|c|}{\textbf{Number of iterations}}      & \multicolumn{1}{c|}{\textbf{1}} & \multicolumn{1}{c|}{\textbf{2}} \\ \hline
\multicolumn{1}{|c|}{\textbf{Total cost deviation (\%)}} & \multicolumn{1}{c|}{0.044}       & \multicolumn{1}{c|}{8.07e-8}       \\ \hline
\multicolumn{1}{|c|}{\textbf{Number of vertices}}        & \multicolumn{1}{c|}{139}       & \multicolumn{1}{c|}{24046}      \\ \hline
\multicolumn{1}{|c|}{\textbf{Time (s)}}                  & \multicolumn{1}{l|}{1.31}        & \multicolumn{1}{l|}{403.74}        \\ \hline
\textbf{}                                                & \multicolumn{1}{l}{}            & \multicolumn{1}{l}{}           
\end{tabular}
\end{table}

\section{Conclusion}
This paper proposes a projection-based aggregation method for distribution networks with DERs, which comprehensively considers various factors such as network security constraints, temporal coupling constraints, and DER cost characteristics. Based on parameters such as the topology of the distribution network, the location of the DERs within the network, output limitations and cost information of DERs, a high-dimensional vertex-type convex hull is calculated as an aggregation model of the DERs cluster. The convex hull fully characterizes the gate power and cost characteristics of the distribution network with DERs, enabling the DERs cluster to participate in the ED of the smart grid as a whole. The case study shows that the results of the centralized ED of distribution networks with DERs are the same as those participating in the ED of the smart grid after aggregation, indicating the effectiveness of the aggregation method and model. 

% \begin{refcontext}[sorting = none]
% \printbibliography
% \end{refcontext}

\bibliographystyle{IEEEtran}
% \bibliography{EI.bib}

\end{document}